\documentstyle[12pt]{article}
\begin{document}

\begin{center}
{\Large \bf Axial Anomaly through Analytic Regularization}
\end{center}
\vspace{1.2cm}
\begin{center}
L. A. Manzoni and B. M. Pimentel\linebreak[1] \linebreak[1]
Instituto de F\'{\i}sica Te\'{o}rica\linebreak[1]
Universidade Estadual Paulista\linebreak[1]
Rua Pamplona, 145\linebreak[1]
01405-900 - S\~ao Paulo, S.P. \linebreak[1]
Brazil\linebreak[1]
\linebreak[1]
and
\linebreak[1]
\linebreak[1]
J. L. Tomazelli\linebreak[1] \linebreak[1]
Departamento de F\'{\i}sica e Qu\'{\i}mica - Faculdade de Engenharia\linebreak[1]
Universidade Estadual Paulista - Campus de Guaratinguet\'a\linebreak[1]
Av. Dr. Ariberto Pereira da Cunha, 333\linebreak[1]
12500-000 - Guaratinguet\'a, S.P. \linebreak[1]
Brazil
\end{center}

\vspace{2.2cm}
\begin{abstract}
In this work we consider the 2-point Green's functions in (1+1) dimensional quantum electrodynamics and show that the correct implementation of analytic regularization gives a gauge invariant result for the vacuum polarization amplitude and the correct coefficient for the axial anomaly. 
\end{abstract}

\vspace{0.5cm}


keywords: Analytic regularization; Ward identity; axial anomaly; vacuum polarization 

\pagebreak

It is well established that some quantum field theories do not preserve all of the symmetries which are present in the corresponding classical theories. As a consequence, we are faced with anomalous Ward identities when we set up relations among the Green's functions of such theories. An example is the famous axial anomaly in a theory with gauge coupled fermions. 

Another well known fact is that, in the derivation of the anomaly, the result may depend on the regularization method employed and the possible regularization schemes are restricted due to the requirement of gauge symmetry.

A possible method to handle with the ultraviolet divergencies is the analytic regularization, introduced by Bollini {\it et al.}\cite{bgg}. Essentially, this technique consists in replacing the denominator of a propagator $(p^2-m^2+i\varepsilon )$, in a divergent amplitude, by $(p^2-m^2+i\varepsilon )^{1+\alpha}$ and take the $\alpha$ parameter large enough to keep the corresponding integrals finite. Then, the resulting amplitude is an analytic function of $\alpha$, except for isolated singularities which appear as poles in $\Gamma$-functions. The regularized expression is obtained by considering the finite part of the Laurent series at $\alpha =0$.

Despite of the algebraic advantages of this method, it has run into oblivion because in its original formulation the gauge symmetry is spoiled when we consider fermion loops. Three decades ago Breitenlohner and Mitter\cite{bmi} proposed an extension of the analytic regularization technique in which the fermion loops are regularized as a whole, much in the same spirit as in Pauli-Villars regularization\cite{pvi}, in such way that the gauge invariance is preserved. Hence, the analytic regularization still applies and, in special, permit us to tackle anomalous amplitudes in a simple way\cite{bmi2}.

 It is our pourpose in this letter to show, by a direct calculation of the relevant Feynman diagrams, that applying naive analytic regularization to (1+1) dimensional quantum electrodynamics (QED$_2$) both the vector and the axial Ward identities are broken. But, if  we take into account the modifications sugested in \cite{bmi} we recover the gauge invariance.

The Lagrangian density for QED$_2$ is given by
\begin{equation}
{\cal L}=\overline{\psi}\left(i\partial\!\!\!\slash - eA\!\!\!\slash -m\right)\psi -\frac{1}{4}F_{\mu\nu}F^{\mu\nu} , 
\label{lag}
\end{equation}

\noindent
where $F_{\mu\nu}$ stands for the usual field strength tensor and $\mu ,\nu = 0,1$.

In two-dimensional space-time the algebra for Dirac gamma matrices is realized using the Pauli matrices
\begin{equation}
\gamma_0=-\sigma_2 \; ,\hspace{0.8cm}\gamma_1=-i\sigma_1 \; ,\hspace{0.8cm}\gamma_5=\gamma_0\gamma_1=\sigma_3 \; ,
\end{equation}

\noindent
with $g_{\mu\nu}={\rm diag}(1,-1)$.

We are concerned with the two-point functions relevant for the axial anomaly in QED$_2$
\begin{eqnarray}
T_{\mu\nu}(x-y)&=& \langle 0 | T\{ V_{\mu}(x) V_{\nu}(y)\}|0\rangle \; , \\ \nonumber \\
T_{\mu\nu}^5(x-y)&=& \langle 0 | T\{ V_{\mu}(x) A_{\nu}(y)\}|0\rangle \; , \\ \nonumber \\
P_{\mu}^5(x-y)&=& \langle 0 | T\{ V_{\mu}(x) P(y)\}|0\rangle \; , 
\end{eqnarray}

\noindent
where $V_{\mu}$, $A_{\mu}$ and $P$ are the vector, axial-vector and pseudoscalar currents, respectively,
\begin{eqnarray}
V_{\mu}(x)&=&\overline{\psi}(x)\gamma_{\mu}\psi (x) \; , \\ \nonumber \\
A_{\mu}(x)&=&\overline{\psi}(x)\gamma_{\mu}\gamma_5\psi (x) \; , \\ \nonumber \\
P(x)&=&\overline{\psi}(x)\gamma_5\psi (x) \; .
\end{eqnarray}

\noindent
The axial anomaly comes from $T_{\mu\nu}^5$. Since, in two dimensions, $\gamma^{\mu}\gamma_5 = \epsilon^{\mu\nu}\gamma_{\nu}$, we have the relation
\begin{equation}
T_{\mu\nu}^5(x-y)=\epsilon_{\nu\beta}T_{\mu}^{\;\;\beta}(x-y).
\label{vv-av}
\end{equation}

\noindent
Therefore, in second-order perturbation theory, we must calculate just the fermion loop containing two vector currents, that is,  the vacuum polarization tensor. In addition, we will need the fermion loop containing the vector current and the pseudoscalar current $P_{\mu}^5$. Let us start considering the vector-vector tensor in momentum space
\begin{equation}
T_{\mu\nu}(k)=i\int \frac{d^2p}{(2\pi )^2} \frac{{\rm tr}\{ \gamma_{\mu}\left[ (p\!\!\!\slash +k\!\!\!\slash) + m\right]\gamma_{\nu}(p\!\!\!\slash +m)\} }{\left[ (p+k)^2 -m^2\right] (p^2 - m^2 ) }.
\label{vv}
\end{equation} 

\noindent
This expression is logarithmically divergent by power counting and we must regularize it to perform the calculations. We will begin by considering the naive analytic regularization technique, which may be implemented by replacing one of the fermion propagators in (\ref{vv}) by
\begin{equation}
S^{(\alpha )}(p,m)=m^{2\alpha}f(\alpha )  \frac{(p\!\!\!\slash +m)}{(p^2-m^2+i\varepsilon )^{1+\alpha }} , \label{propmod}
\end{equation}

\noindent
where the $m^{2\alpha}$ term is to ensure that the correct dimension of mass will be maintained; $f(\alpha )$ is a regular function of $\alpha$ with $f(0)=1$.

Due to the above regularization prescription, a number of  arbitrary and finite constants may result from the regularized amplitude. These constants are the coefficients of the lowest order terms in the expansion of $f(\alpha )$ in Taylor series. Thus, we have a certain degree of arbitrariness related to $f(\alpha  )$.

The regularized vacuum polarization tensor is given by
\begin{equation}
\tilde{T}_{\mu\nu}^{\;(\alpha)}(k)=im^{2\alpha}f(\alpha )\int \frac{d^2p}{(2\pi )^2} \frac{{\rm tr}\{ \gamma_{\mu}\left[ (p\!\!\!\slash +k\!\!\!\slash) + m\right]\gamma_{\nu}(p\!\!\!\slash +m)\} }{\left[ (p+k)^2 -m^2\right] (p^2 - m^2 )^{1+\alpha} } ,
\label{tt}
\end{equation} 

\noindent
where the tilde is to indicate the naively implemented analytic regularization. Now, taking the $\alpha$ parameter large enough to keep the integral convergent, we can make use of the symmetric integration and perform shifts in the integration variable with confidence.

We proceed by introducing the Feynman parametrization in (\ref{tt}). Then, taking the trace and performing the momentum integration, we can write down $\tilde{T}_{\mu\nu}^{(\alpha )}$ as an integral in the auxiliary variable $\xi$: 
\begin{eqnarray}
\tilde{T}_{\mu\nu}^{\;(\alpha)}(k)&=&-\frac{(-1)^{\alpha}}{4\pi }m^{2\alpha}f(\alpha )\int_0^1 d\xi  \frac{(1-\xi )^{\alpha}}{(M^2)^{1+\alpha}} \nonumber \\ \nonumber \\
&\times& \{ -4\xi (1-\xi )\left[ k_{\mu}k_{\nu} - g_{\mu\nu}k^2 \right] + 2g_{\mu\nu}M^2 \} \label{ttil} ,
\end{eqnarray} 

\noindent
where we have defined
\begin{equation}
M^2=m^2-\xi (1-\xi )k^2 .
\end{equation}

\noindent
The expression (\ref{ttil}) is finite in $\alpha$, so that it is safe to put $\alpha =0$ to obtain 
\begin{equation}
\tilde{T}_{\mu\nu}(k) = \left( \frac{k_{\mu}k_{\nu}}{k^2} - g_{\mu\nu}\right) \Pi (k^2) -\frac{1}{2\pi}g_{\mu\nu} ,
\label{naivevv}
\end{equation}

\noindent
with
\begin{eqnarray}
\Pi (k^2) &=& -\frac{1}{\pi}+\frac{m^2}{\pi}\int_0^1 d\xi \frac{1}{m^2-\xi (1-\xi )k^2} \nonumber \\ \nonumber \\
&=&-\frac{1}{\pi}\left[ 1 + \frac{2m^2}{k^2\sqrt{1-\frac{4m^2}{k^2}}}\ln \left( \frac{\sqrt{1-\frac{4m^2}{k^2}}+1}{\sqrt{1-\frac{4m^2}{k^2}}-1}\right) \right]   .   \label{pi}
\end{eqnarray}

\noindent
From (\ref{naivevv}) we see that the analytic regularization technique does not render a gauge invariant vacuum polarization tensor in 2 dimensions, if naively implemented. The same conclusion is obtained when this technique is applied to calculate the vacuum polarization tensor in 4 dimensions\cite{bmi}. As already pointed in ref. \cite{bmi}, the reason for this is that the modified propagator (\ref{propmod}) does not fulfil Ward's identity.

However, it is worth noting that the naive analytic regularization technique can be applied to (2+1)-dimensional gauge theories to one loop\cite{hmo}. The results obtained in \cite{hmo} are in accordance with those obtained considering the construction of ref. \cite{bmi} (see \cite{pst}).

Let us now consider the fermion loop containing a vector and a pseudoscalar current, given by
\begin{equation}
P_{\mu}^5(k)=i\int \frac{d^2p}{(2\pi )^2} \frac{{\rm tr}\{ \gamma_{\mu}\left[ (p\!\!\!\slash +k\!\!\!\slash) + m\right]\gamma_5(p\!\!\!\slash +m)\} }{\left[ (p+k)^2 -m^2\right] (p^2 - m^2 ) }.
\end{equation} 

\noindent
Noticing that the trace in this expression is equal to $-2m\epsilon_{\mu\nu}k^{\nu}$, we verify that $P_{\mu}^5$ is ultraviolet finite and can be evaluated to give
\begin{equation}
P_{\mu}^5(k)= -\frac{m}{\pi}\epsilon_{\mu\nu}k^{\nu}\frac{1}{k^2\sqrt{1-\frac{4m^2}{k^2}}}\ln \left( \frac{\sqrt{1-\frac{4m^2}{k^2}}+1}{\sqrt{1-\frac{4m^2}{k^2}}-1}\right) .
\label{pv}
\end{equation} 

\noindent
Then, from (\ref{naivevv}), (\ref{pi}) and (\ref{pv}), we get
\begin{eqnarray}
k^{\mu}\tilde{T}_{\mu\nu}^5(k)&=&-\frac{1}{2\pi}\epsilon_{\nu\mu}k^{\mu} , \\ \nonumber \\
k^{\nu}\tilde{T}_{\mu\nu}^5(k)&=&2mP_{\mu}^5(k)-\frac{1}{2\pi}\epsilon_{\mu\nu}k^{\nu}.
\end{eqnarray}

\noindent
So, the anomaly is equally distributed between the vector and the axial-vector Ward identities. This result is in accordance with that obtained in \cite{abh}, using a naive kind of regularization where the singular term is just omitted. However, this is not satisfactory because we would like to preserve gauge invariance.

We now perform the same calculation using the analytic loop procedure developed in \cite{bmi}. Here we only consider the vacuum polarization (for a general treatment see the original work\cite{bmi}).

The first step to implement the analytic loop regularization to the vacuum polarization tensor is to rewrite it using the Feynman parametrization as 
\begin{equation}
T_{\mu\nu}(k)=i\int \frac{d^2p}{(2\pi )^2}\int_0^1d\xi_1\int_0^1d\xi_2\delta (1-\xi_1-\xi_2) \left[\frac{ P(m)}{\left[ z -m^2+i\varepsilon\right]^2 }\right]_{z=\sum_{j=1}^2\xi_jp_j^2},
\end{equation} 

\noindent
where the polinomial $P(m)$ is defined as
\begin{equation}
P(m)\equiv {\rm tr}\{ \gamma_{\mu}( {p\!\!\!\slash}_1 + m)\gamma_{\nu}({p\!\!\!\slash}_2 +m)\}, 
\end{equation}

\noindent
with $p_1\equiv p+k$ and $p_2 \equiv p$.

Then, introducing another polinomial
\begin{equation}
Q(\kappa )\equiv \frac{1}{2\kappa}\left[ (m+\kappa ) P(\kappa ) - (m- \kappa ) P(-\kappa )\right] ,
\end{equation}

\noindent
such as $P(m)= Q(\kappa^2)|_{\kappa^2=m^2}$, we obtain
\begin{eqnarray}
T_{\mu\nu}(k)&=&-i\int \frac{d^2p}{(2\pi )^2}\int_0^1d\xi_1\int_0^1d\xi_2\; \delta (1-\xi_1-\xi_2)  \nonumber \\ \nonumber \\
&\times&\left[\frac{d}{dz}\left(\frac{ S(\kappa )P(\kappa )-S(-\kappa )P(-\kappa )}{2\kappa}\right)_{\kappa^2=z}\right]_{z=\sum\xi_jp_j^2} , \label{tgau}
\end{eqnarray} 

\noindent
where $S(p)$ stands for the fermion propagator. Now, the prescription consists in replacing the fermion propagator in (\ref{tgau}) by its modified form (\ref{propmod}), so that 
\begin{equation}
T_{\mu\nu}^{(\alpha )}(k)=-i\int \frac{d^2p}{(2\pi )^2}\int_0^1d\xi_1\int_0^1d\xi_2\;\delta (1-\xi_1-\xi_2) \left[\frac{d}{dz}F^{(\alpha )}(z)\right]_{z=\sum\xi_jp_j^2} ,
\label{vvreggauge}
\end{equation} 

\noindent
where
\begin{equation}
F^{(\alpha )}(z)\equiv \left(\frac{ S^{(\alpha )}(\kappa ,m)P(\kappa )-S^{(\alpha )}(-\kappa ,m)P(-\kappa )}{2\kappa}\right)_{\kappa^2=z}.
\label{F}
\end{equation}

\noindent
So, the expression for the vector-vector tensor is convergent for $\alpha $ sufficiently large. Substituting (\ref{propmod}) in (\ref{F}) we obtain
\begin{eqnarray}
\frac{d}{dz}F^{(\alpha )}(z)&=&-\frac{(-1)^{-\alpha}2(1+\alpha )m^{2\alpha}f(\alpha )}{(m^2-z-i\varepsilon )^{2+\alpha}}\left[ g_{\mu\nu}(z-p_1\cdot p_2)+ p_{1\mu}p_{2\nu}+p_{2\mu}p_{1\nu}\right]  \nonumber \\ \nonumber \\
&-&\frac{(-1)^{-\alpha}2m^{2\alpha}f(\alpha )}{(m^2-z-i\varepsilon )^{1+\alpha}}\; g_{\mu\nu}\; .   \label{dF}
\end{eqnarray}

\noindent
Inserting (\ref{dF}) in (\ref{vvreggauge}) and integrating with respect the auxiliary parameter $\xi_2$ we obtain, after performing the shift $p\rightarrow p -\xi_1 k$ in the momentum variable,
\begin{eqnarray}
T_{\mu\nu}^{(\alpha )}(k)\!&=&\!(-1)^{-\alpha}2im^{2\alpha}f(\alpha )\int_0^1d\xi \int \frac{d^2p}{(2\pi )^2}\left[ \frac{1+\alpha}{(M^2-p^2-i\varepsilon )^{2+\alpha}}\{ 2\xi (\xi -1) \right.  \nonumber \\ \nonumber \\
&\times&  \left. \! (k_{\mu}k_{\nu}-k^2g_{\mu\nu})  +2k_{\mu}k_{\nu}\} + \frac{g_{\mu\nu}}{(M^2-p^2-i\varepsilon )^{1+\alpha}} \right]  . \label{vvrg1}
\end{eqnarray} 

\noindent
A glance in the above expression shows that the last two terms in the square bracket break the gauge invariance. Performing the momentum integration, we get for these terms
\begin{eqnarray}
_{GB1}\Pi^{(\alpha )}_{\mu\nu}&=&4im^{2\alpha}f(\alpha )(-1)^{-\alpha}(1+\alpha )\int_0^1 d\xi \int  \frac{d^2p}{(2\pi )^2} \frac{k_{\mu}k_{\nu}}{(M^2-p^2-i\varepsilon )^{2+\alpha}} \nonumber \\ \nonumber \\
&=&g_{\mu\nu}\frac{(-1)^{\alpha}}{2\pi}f(\alpha )(1+\alpha )\frac{\Gamma (\alpha )}{\Gamma (2+\alpha )}\int_0^1d\xi \frac{1}{\left[ M^2-i\varepsilon \right]^{\alpha}} ,
\end{eqnarray}

\noindent
and
\begin{eqnarray}
_{GB2}\Pi^{(\alpha )}_{\mu\nu}&=&g_{\mu\nu}2im^{2\alpha}(-1)^{-\alpha}f(\alpha )\int_0^1 d\xi \int  \frac{d^2p}{(2\pi )^2} \frac{1}{(M^2-p^2-i\varepsilon )^{1+\alpha}} \nonumber \\ \nonumber \\
&=&-g_{\mu\nu}\frac{(-1)^{\alpha}}{2\pi}f(\alpha )\frac{\Gamma (\alpha )}{\Gamma (1+\alpha )}\int_0^1d\xi \frac{1}{\left[ M^2-i\varepsilon \right]^{\alpha}} .
\end{eqnarray}

\noindent
Therefore, they cancel each other. So that the only remaining term in (\ref{vvrg1}) is the transverse one, which is finite when $\alpha \rightarrow 0$ and equals to $\Pi (k^2)$ in eq. (\ref{pi}).

Then, the analytic loop regularization gives a gauge invariant expression to the vacuum polarization tensor in two dimensions,
\begin{equation}
T_{\mu\nu}(k) = \left( \frac{k_{\mu}k_{\nu}}{k^2} - g_{\mu\nu}\right) \Pi (k^2)  .
\end{equation}

\noindent
From this expression and identity (\ref{vv-av}) we obtain $T_{\mu\nu}^5 (k)$, which satisfies
\begin{eqnarray}
k^{\mu}T_{\mu\nu}^5&=&0  ,\\ 
k^{\nu}T_{\mu\nu}^5&=&2mP_{\mu}^5(k)+\frac{1}{\pi}\epsilon_{\nu\mu}k^{\nu} ,
\end{eqnarray}

\noindent
so that the only anomalous Ward identity is the axial one, and the coefficient of the anomaly is just the photon mass, as can be seen from (\ref{pi}). This result is in accordance with that obtained using dimensional and Pauli-Villars regularizations (see ref. \cite{abh}, for example).

In conclusion, we have considered the 2-point functions in QED$_2$ and showed that, using a naively implemented analytic regularization, both the axial and vector Ward identities are spoiled. However, considering the appropriate analytic loop construct\cite{bmi} the gauge symmetry is preserved while the axial current carries the anomaly.

In view of this, we may ask if it is possible to implement analytic regularization in such a way that chiral symmetry is preserved\cite{jac}. This is object of future investigations.

\vspace{1.5cm}
{\bf {\large Acknowledgements}}
\vspace{0.5cm}

L. A. M. is supported by Conselho Nacional de Desenvolvimento Cient\'{\i}fico e Tecnol\'{o}gico (CNPq); B. M. P. and J. L. T. are partially supported by CNPq.

\pagebreak

\end{document}